\documentclass[a4paper,11pt]{article}
\usepackage{jinstpub} 

\makeatletter
\def\@fpheader{}
\makeatother

\usepackage[colorinlistoftodos]{todonotes}
\usepackage{lineno}
\usepackage{float}
\usepackage{amsmath}
\usepackage{subcaption}
\usepackage{caption}
\usepackage{graphicx} 
\usepackage{verbatim} 

\title{\boldmath Enhancing Water Cherenkov Detector Arrays through multiPMT Modules}

\author[a,b]{D. Ambrosino,}
\author[a,b]{R. Colalillo,}
\author[c,b]{V. M. Grieco$^{1}$, \note{Corresponding author.}}
\author[a,b]{F. Guarino,}
\author[a,b]{L. Lavitola,}
\author[a,b]{F.~Sansone,}
\author[a,b]{M. Tambone,}
\author[a,b]{L. Valore,}
\author[a,b]{M. Waqas}

\affiliation[a]{Università degli Studi di Napoli "Federico II",\\Complesso Universitario di Monte Sant'Angelo, Dipartimento di Fisica, via Cintia 80126 Napoli, Italy}
\affiliation[b]{Istituto Nazionale di Fisica Nucleare, Sezione di Napoli,\\Complesso Universitario di Monte Sant'Angelo, Dipartimento di Fisica, via Cintia 80126 Napoli, Italy}
\affiliation[c]{Scuola Superiore Meridionale, Via Mezzocannone 4, 80138 Napoli, Italy}

\emailAdd{v.grieco@ssmeridionale.it}

\abstract{Water Cherenkov Detectors (WCDs) are pivotal in various scientific fields, including neutrino physics, gamma-ray astronomy, and cosmic-ray research. The detection sensitivity and precision of these detectors crucially rely on photomultiplier tubes (PMTs) to capture Cherenkov radiation produced by charged particles moving faster than the speed of light in water. In recent years, employing multiPMT modules 
has emerged as a promising strategy to enhance large volume water and ice Cherenkov detector performance. In this work we explore 
the use of a multiPMT module in 
small WCD units, arranged in arrays as typically used 
to detect Extensive Air Showers (EAS). We outline a possible configuration and present the advantages it can offer 
for data analysis, as demonstrated through dedicated simulations. We investigate the potential of multiPMTs in capturing the features of the Cherenkov light distribution originated by single muons and discuss its possible application for muon tagging in WCD arrays.}


\begin{document}
\maketitle
\flushbottom

\section{Introduction}
In the framework of multimessenger astronomy, the detection of high-energy gamma rays, neutrinos, and cosmic rays requires specialized techniques and instruments suited to the characteristics and energy of the particles \cite{Funk_2015,SILIANG2021281, TAJIMA200712}. Ground-based detectors play a central role in this effort at high energies. By capturing the cascades of secondary particles produced in the atmosphere, they allow the characteristics of the primary particle to be inferred through the reconstruction of EAS properties \cite{Errando_2023}. Among the available techniques, arrays of WCD units have been proven particularly effective in ground-based observatories \cite{spiering2023cherenkovdetectorsastroparticlephysics}. In these setups, charged particles moving through water faster than the speed of light in the medium emit Cherenkov radiation which is then detected by photosensors. Experiments such as the Pierre Auger Observatory~\cite{Allekotte_2008, PAO_SD}, the High-Altitude Water Cherenkov (HAWC) Observatory \cite{Abeysekara_2023, pretz2015highlightshighaltitudewater}, and the Large High Altitude Air Shower Observatory (LHAASO) \cite{WCDA_LHAASO, LHAASO_wcda2, Chen:2020xw} rely on this methodology to investigate cosmic rays and gamma rays. A similar technique is also used in neutrino astronomy, where large volume detectors such as Super-Kamiokande \cite{Walter_2008} and IceCube \cite{Halzen_2010} record Cherenkov light from charged leptons, the latter being produced by neutrino interactions inside the detector volume.

Traditionally, Cherenkov light in WCD arrays has been recorded with large-area PMTs (photocathode diameter $\ge$ 8-inches) \cite{TRIPATHI20031}. However, limitations such as coverage inefficiencies, reduced field of view, cost per photocathode area, limited dynamic range and low production rate have prompted the exploration of multiPMT configurations using clusters of small PMTs to overcome these challenges. The multiPMT module was proposed for the first time for the KM3NeT experiment with the development of the Digital Optical Module (DOM) \cite{km3} and combines several 3-inch PMTs, power supply and read-out electronics inside a pressure-resistant spherical vessel. This new technology has also been adopted for the Hyper-Kamiokande (Hyper-K) experiment \cite{HyperK} and IceCube-Gen2 \cite{IceCube} with some modifications, with the aim of significantly enhancing detection capabilities, for high-precision neutrino physics and astrophysical observations.

In this work we discuss a concept design for a multiPMT module and investigate its potential application in WCD arrays. After describing the conceptual module, we present the additional information provided by such configuration with respect to traditional single-PMTs, such as signals and time responses from individual channels.
We illustrate, through a dedicated \textsc{Geant4} simulation, the expected response of the photosensor in a simplified case of single muons of known inclinations and injection point. We then move on to explore the multiPMT response in a more realistic scenario of 100 TeV proton-induced EAS. We propose a possible definition of charge asymmetry to chart the secondary particles imprint in the WCD unit and highlight its potentiality for future analyses. Finally, we draw the conclusions of this work by discussing the possible usage of multiPMT systems in WCD units to enhance the muon tagging capabilities of upcoming experiments.

\section{Conceptual module description}
\begin{figure}[t]
	\centering
	\hspace{-1 cm}
	\includegraphics[width=0.6\linewidth, trim={0 0 7cm 0}, clip]{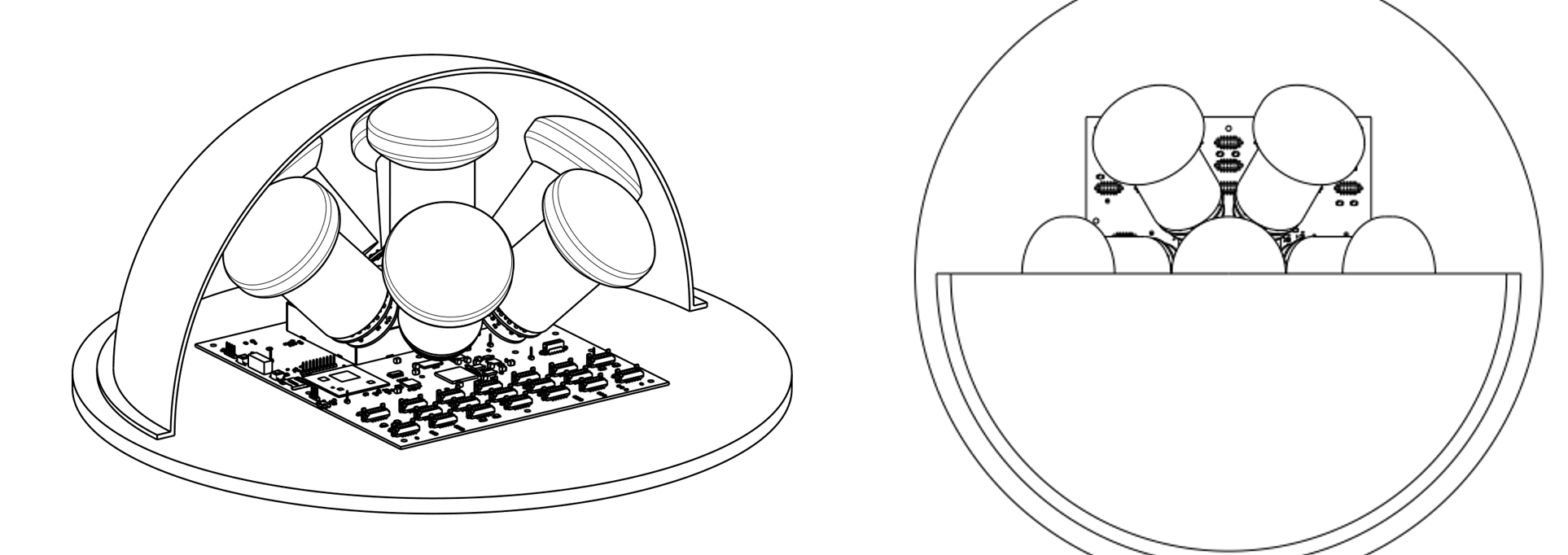}
	\caption{Sketch of the proposed multiPMT module. The semi-spherical vessel hosts seven 3-inch PMTs and the readout electronics within a waterproof enclosure.}
	\label{fig:dome}
\end{figure}

The conceptual multiPMT module we propose in this work is depicted in Fig.\ref{fig:dome}. Its design takes inspiration from the KM3NeT DOM which consists of a spherical module hosting outward-oriented 3-inch PMTs and from the Hyper-K multi-PMT whose design has been revised to be placed on the inner surfaces of the detector.

Our solution consists in a semi-spherical vessel hosting seven 3-inch PMTs: one facing upwards and six tilted by $45^{\circ}$ in zenith and spaced by $60^{\circ}$ in the azimuth. This configuration allows for flexible placement inside a WCD, adaptable to the requirements of different experiments. The number of PMTs was chosen to achieve a total photocathode area comparable to that of a single 8-inch PMT, commonly employed in EAS arrays, ensuring similar photoelectron (PE) collection efficiency. The losses in PE collection efficiency induced by the presence of the vessel can be compensated by optical coupling of the photocatodes and the surfaces of the vessel by means of a gel with a proper refraction index \cite{RUGGERI2025170488} and by the addition of reflectors that increase the light collection area, as already implemented by other experiments \cite{gola2025assemblytestinginstallationmpmt}. Based on current cost estimates, seven 3-inch PMTs provide a lower cost per unit photocathode area than a single 8-inch PMT, making this option more cost-effective.

The use of multiple PMTs offers several advantages over a single larger one, despite the increased system complexity. For example, redundancy in signal detection is provided with a larger number of readout channels, enhancing the robustness of the detector against critical failures, minimizing data loss. Smaller PMTs also offer practical advantages, such as  reduced afterpulsing, improved temporal resolution and  lower sensitivity to external magnetic fields \cite{KOBLESKY201240}. 
Redistributing photons across multiple PMTs alleviates demands on individual readout electronics channels increasing the dynamic range.
Moreover, owing to mass production for KM3NeT, 3-inch Hamamatsu PMTs are extremely reliable and uniform in terms of performance and stability, at a competitive cost \cite{thekm3netcollaboration2025evaluationupgraded3inchhamamatsu}. The semi-spherical vessel provides a water-proof enclosure to co-locate the seven PMTs with the readout electronics. This simplifies installation, testing, and maintenance. Moreover, this solution ensures that the readout electronics and the PMTs would operate in a dry environment at stable operating temperature, as the heat produced by the electronics would be dispersed in the surrounding water.

\label{sec:conceptual-module-description}

\section{Detector simulation and single particle case}

\begin{figure}[t]
	\centering
    \includegraphics[width=1\linewidth, trim = 0 3cm 0 3cm,
  clip]
    {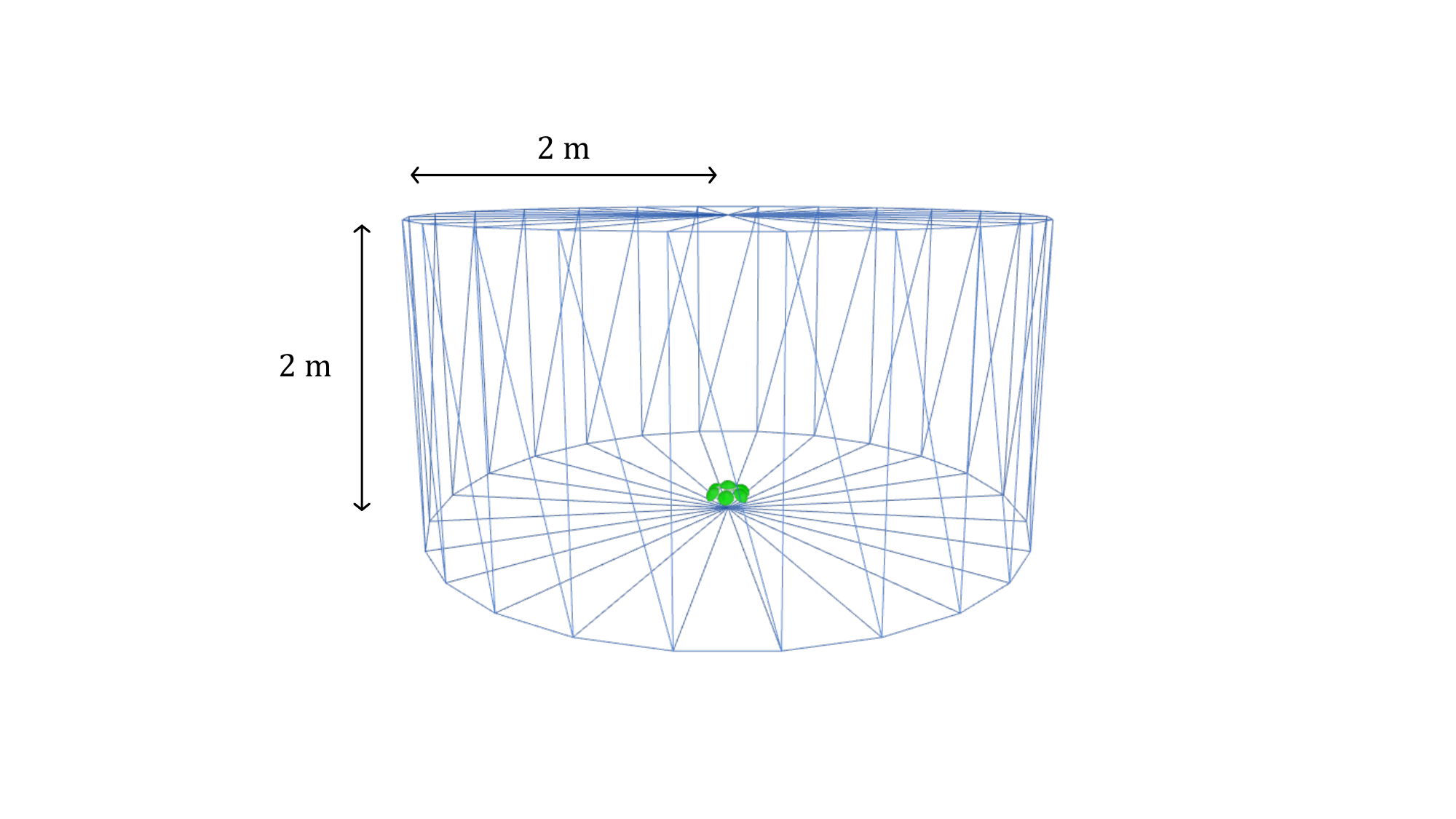}
	\caption{Details of the multiPMT module and of its positioning within the cylindrical water tank ($2$ meters in height and $4$ meters in diameter) simulated with \textsc{Geant4}.}
	\label{fig:simulated}
\end{figure}

We developed the simulation of the multiPMT module using version 11.2 of the \textsc{Geant4} framework \cite{Geant4, Gean4_develop, Geant4_recent} to assess its performance. The simulation incorporates 3-inch PMTs with optical properties such as quantum efficiency, material composition, and sensitive area derived from the Hamamatsu R14374 3-inch PMT datasheet \cite{hamamatsu_3inch}. For the sake of simplicity, the geometrical design of the simulated module focuses solely on the PMT sensitive areas, accurately modeled according to their size and dimensions, while omitting the semi-spherical vessel of the multiPMT which will be included in the actual design. Consequently the PMTs are directly exposed to water. The multiPMT is placed in the bottom center of the WCD unit, a cylindrical water tank that measured 2 meters in height and 2 meters in radius. The internal surface of the tank is non reflective to focus only on direct Cherenkov light. In Fig. \ref{fig:simulated}, we illustrate a rendered picture of the simulated WCD unit adopted in this work.

\begin{figure}[H]
    \centering
    \includegraphics[width=0.48\textwidth]{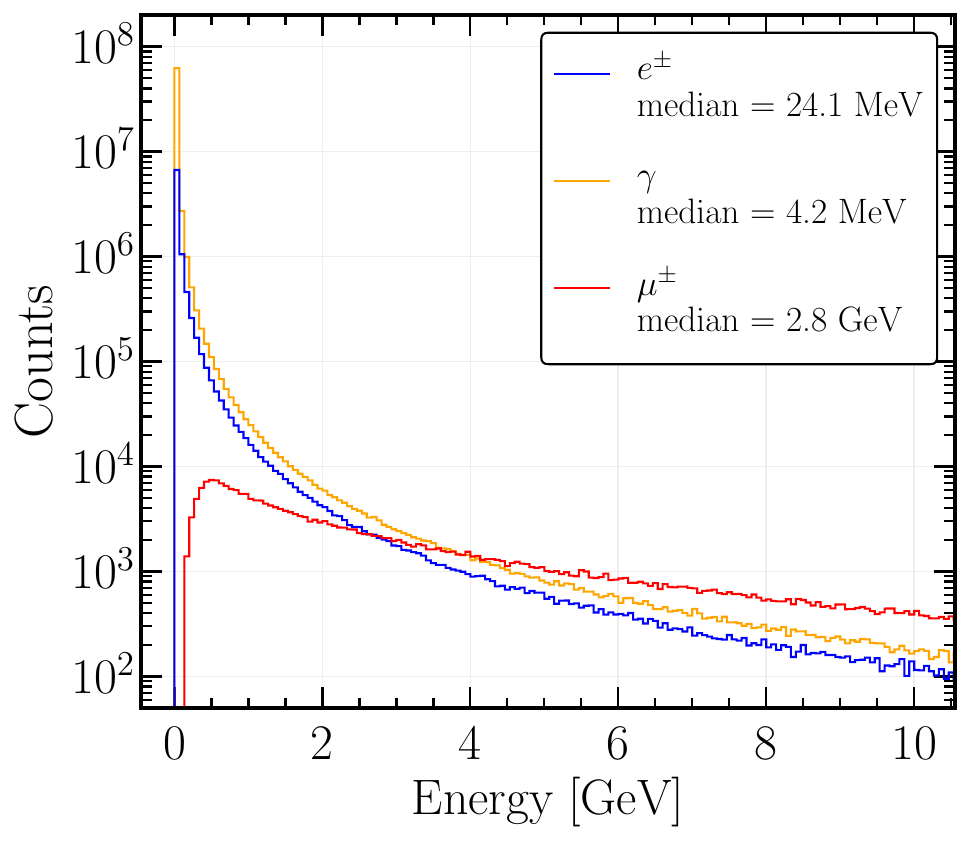}
	\caption{Energy distribution of secondary particles from 200 proton-induced EAS in the energy range between $100 - 110$ $\mathrm{TeV}$ and with axis inclination between $0^{\circ} - 30^{\circ} $.}
 	\label{fig:energy_secondaries}
\end{figure}

In order to study the multiPMT response in the WCD unit we inject single particles with known momentum and point of injection. The energy of such particles is chosen as the median value of the energy spectrum of secondary $\mu^{\pm}$, $\gamma$ and $e^{\pm}$ from 100 - 110 TeV proton-induced EAS at an altitude of 4700 m above sea level, shown in Fig. \ref{fig:energy_secondaries}. We represent the signal obtained solely by single muons at 3 GeV since $\gamma$ and $e^{\pm}$ produce a negligible number of PEs at their respective median energy value. A comprehensive study of all secondary particles will be presented in the next section.

\begin{figure}[p]
    \centering

    \begin{subfigure}[t]{0.48\textwidth}
        \centering
        \includegraphics[width=1\linewidth]{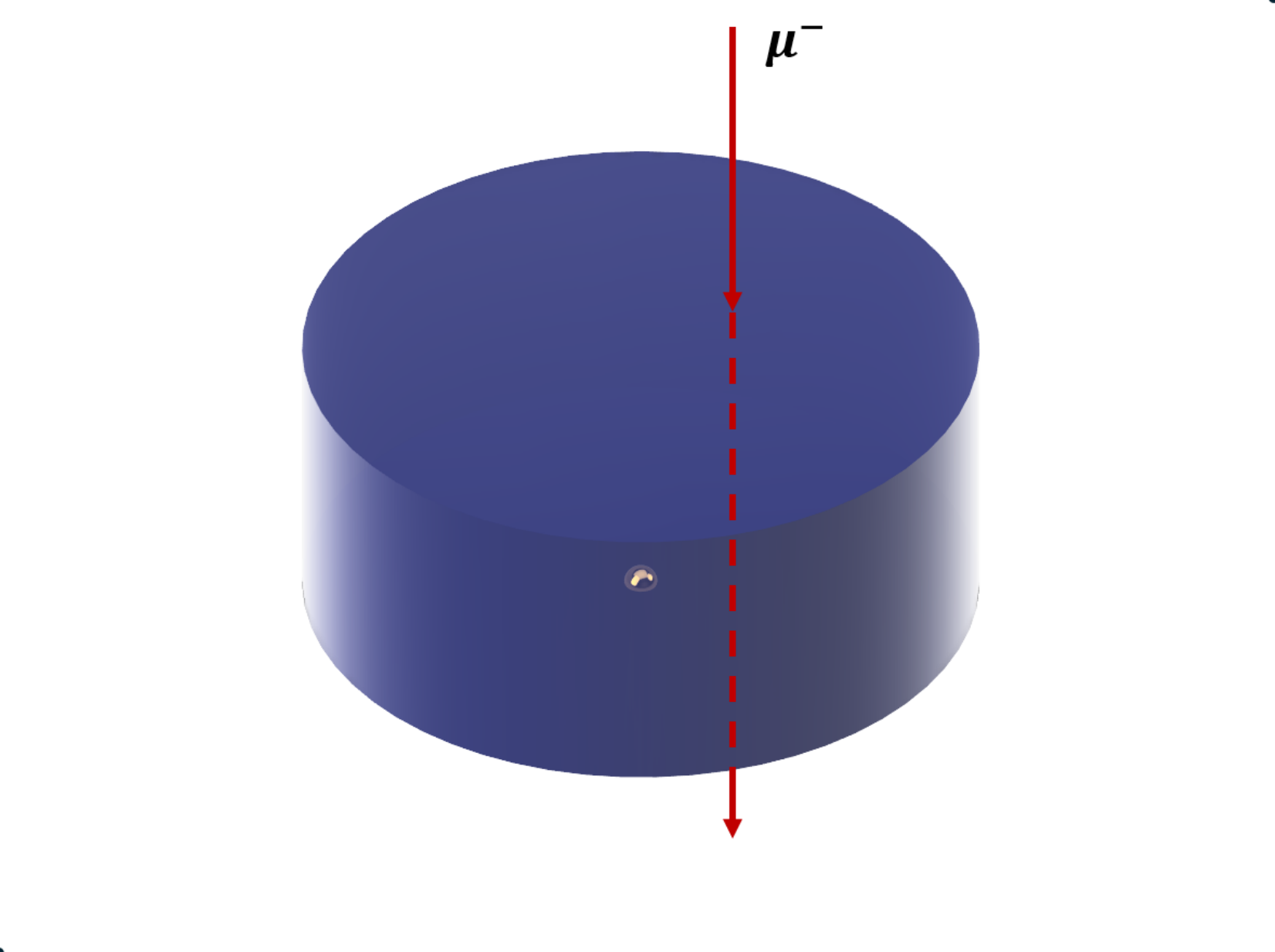}
    \end{subfigure}
    \hfill
    \begin{subfigure}[t]{0.48\textwidth}
        \centering
        \includegraphics[width=0.9\linewidth]{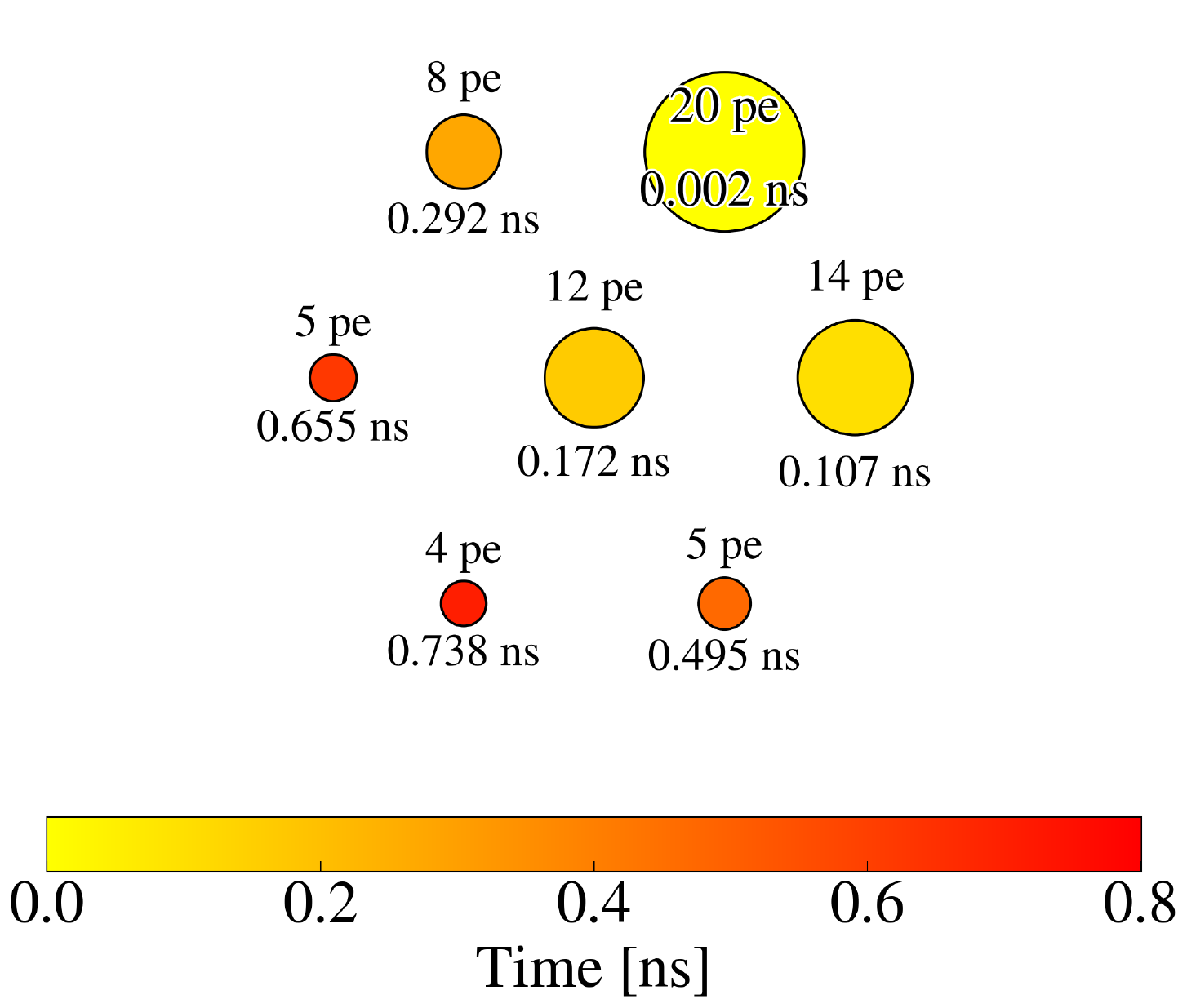}
    \end{subfigure}

    \vspace{0.5cm}

    \begin{subfigure}[t]{0.48\textwidth}
        \centering
        \includegraphics[width=1\linewidth]{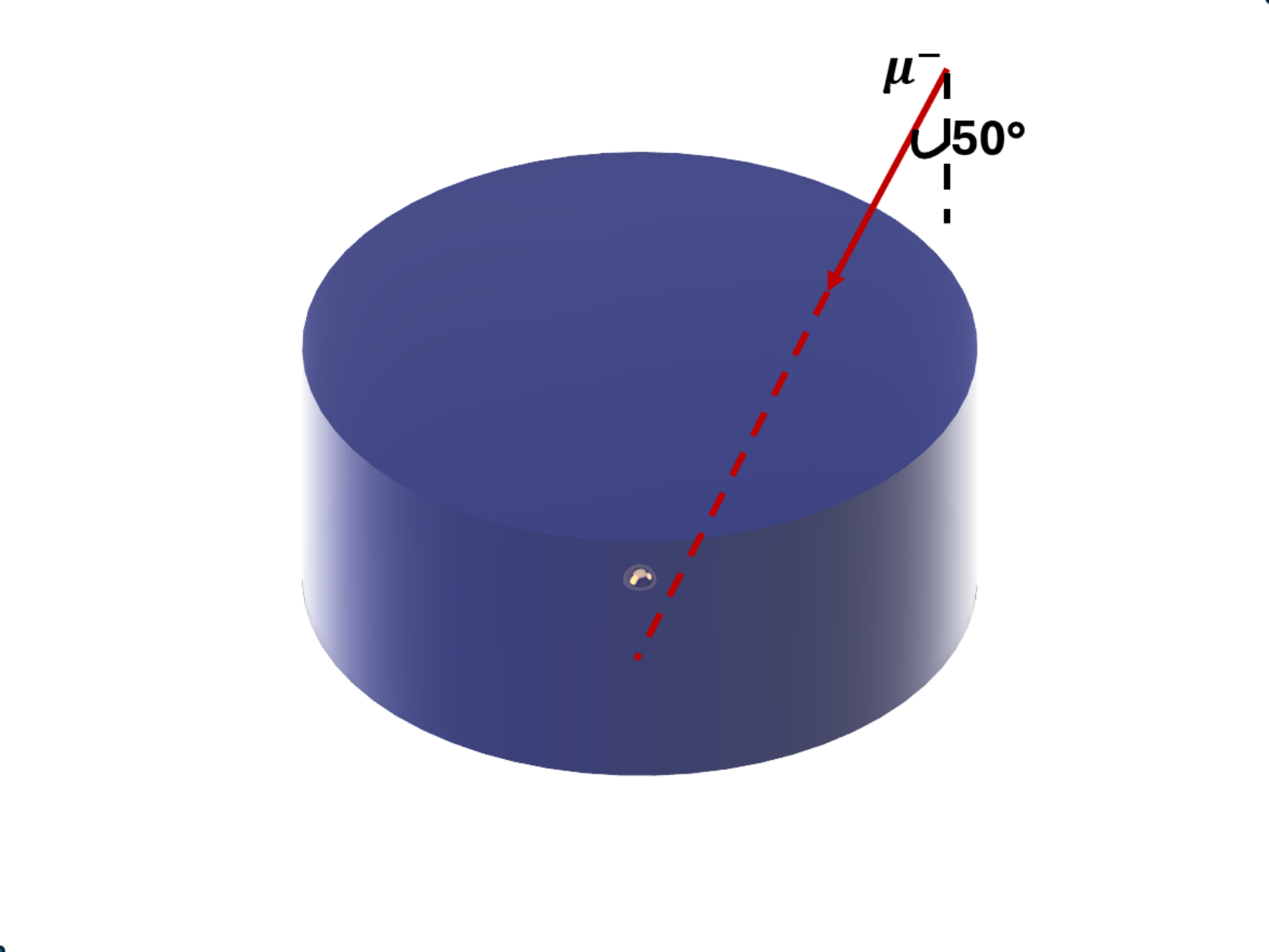}
    \end{subfigure}
    \hfill
    \begin{subfigure}[t]{0.48\textwidth}
        \centering
        \includegraphics[width=0.9\linewidth]{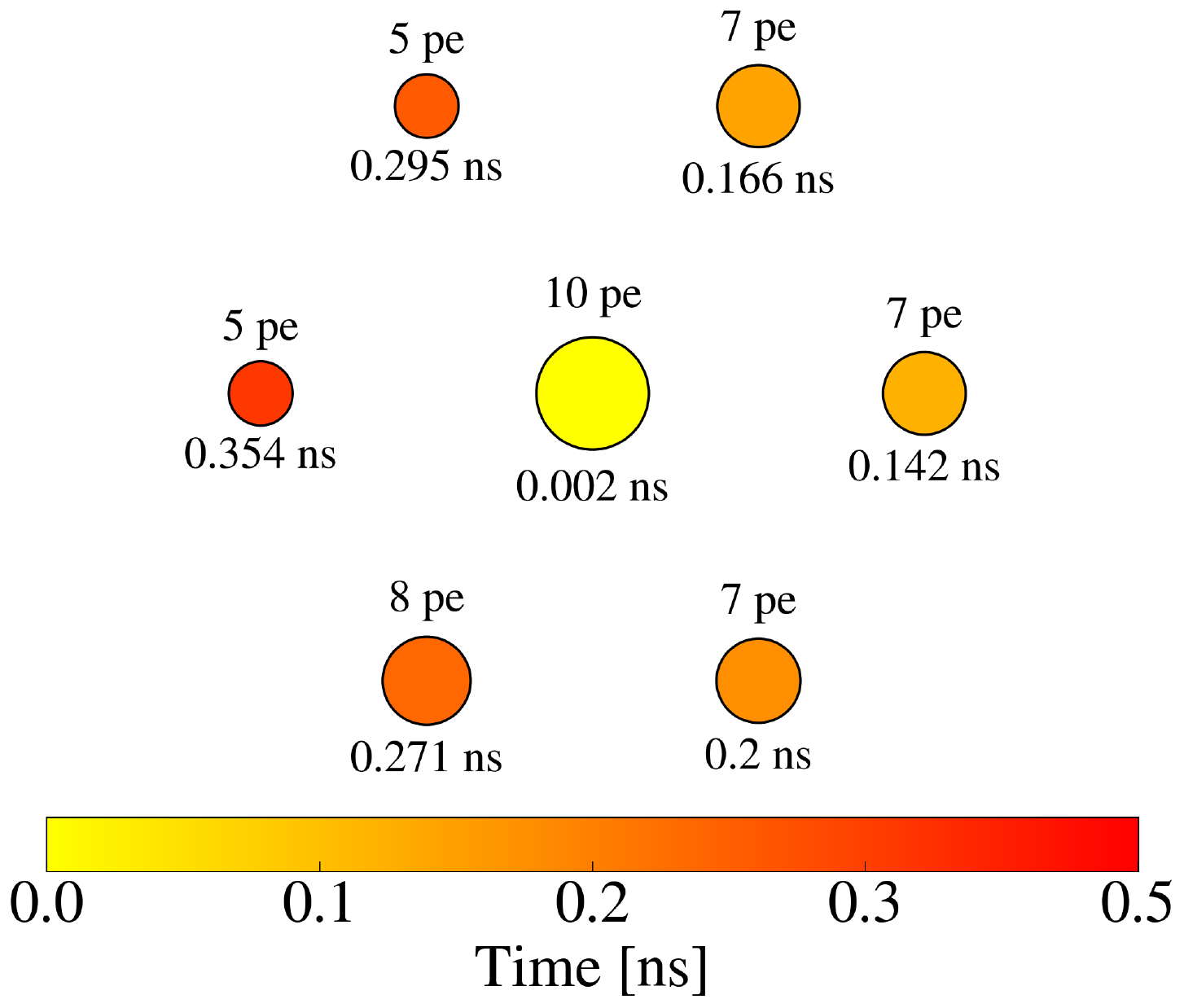}
    \end{subfigure}

    \vspace{0.5cm}

    \begin{subfigure}[t]{0.48\textwidth}
        \centering
        \includegraphics[width=1\linewidth]{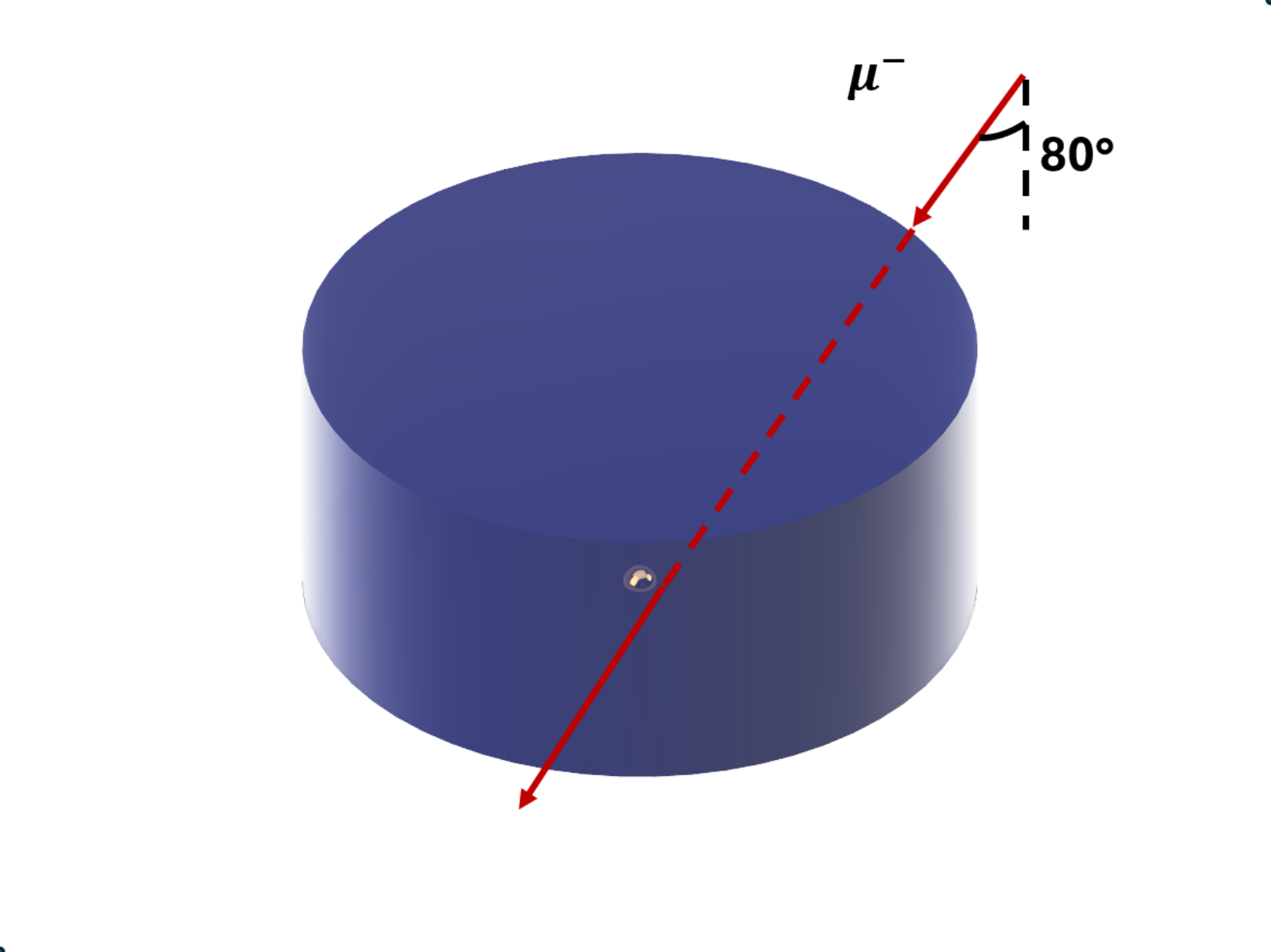}
    \end{subfigure}
    \hfill
    \begin{subfigure}[t]{0.48\textwidth}
        \centering
        \includegraphics[width=0.9\linewidth]{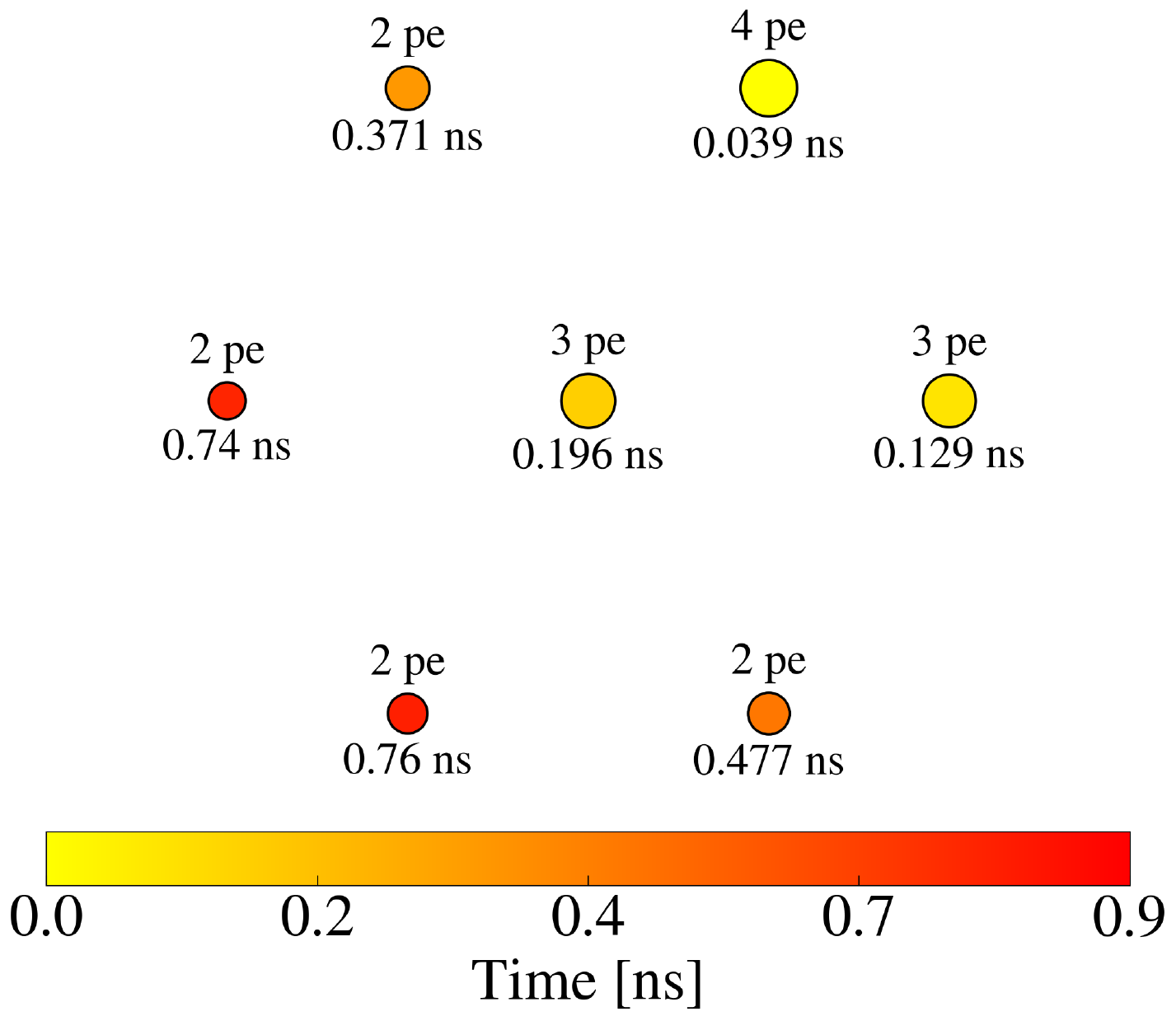}
    \end{subfigure}

    \caption{Simulated \textsc{Geant4} response of a WCD unit equipped with a multiPMT module. Each plot illustrates the average total PE amount per channel and first photon interaction time stamps for a $3$ $\mathrm{GeV}$ single muon of different injection methods simulated 1000 times. The images show respectively the cases of vertical, down-going muons (top image) with an entry position offset with respect to the center, inclined muons traversing the tank from top to bottom with an inclination of $\theta = 50^{\circ}$ (center image), and inclined muons intersecting the WCD on the lateral sides with inclination $\theta = 80^{\circ}$ (bottom image).}
    \label{fig:Augerlike}
\end{figure}

\begin{figure}[h!]
	\centering
    \hspace{2cm}
	\includegraphics[width=0.6\linewidth]{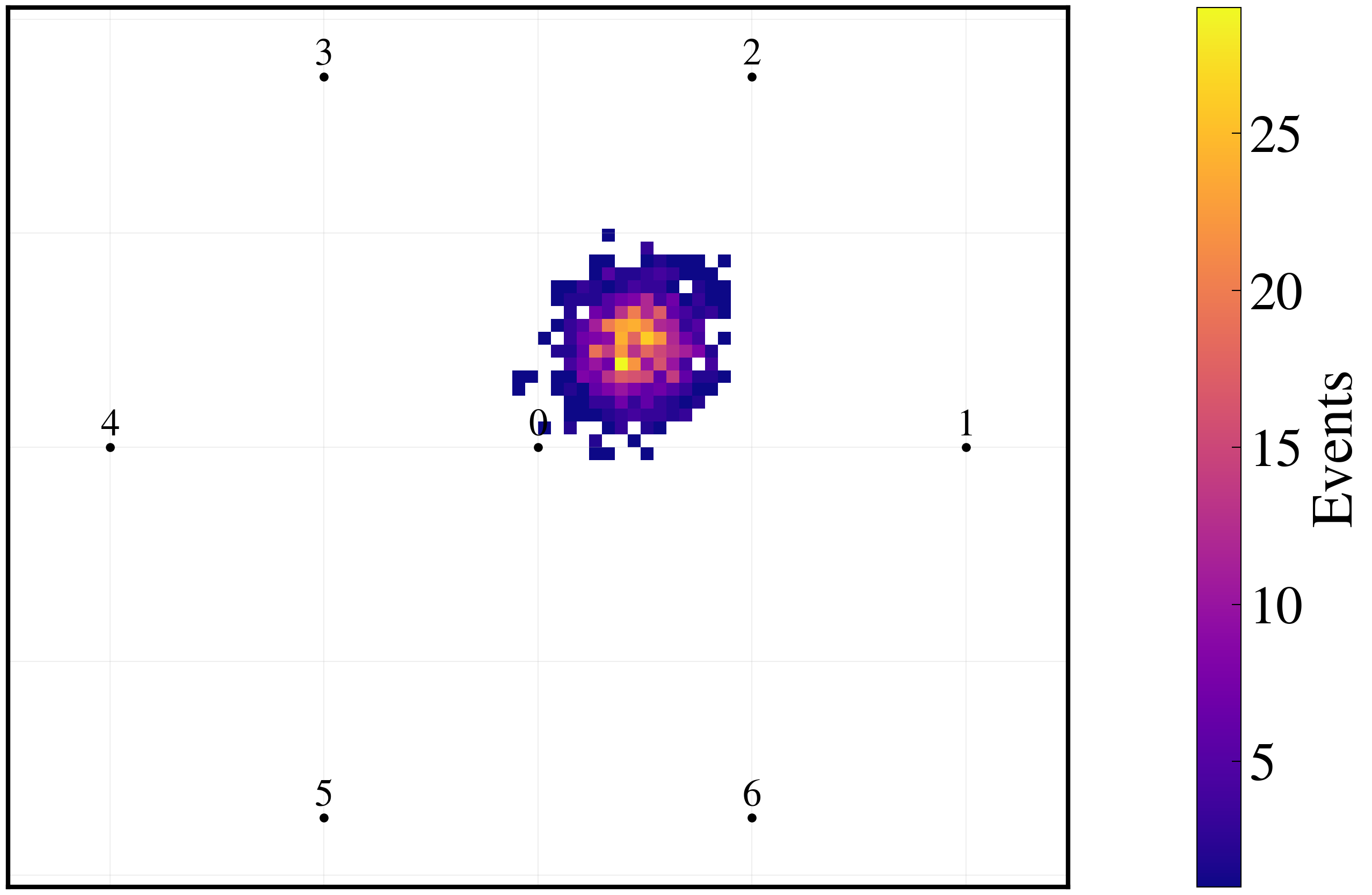}
	\caption{2D histogram of the charge barycenters for 1000 vertical, down-going muons corresponding to the top configuration in Fig. \ref{fig:Augerlike}. The reconstructed barycenter is indicative of the muons off-axis initial position.}
	\label{fig:centroid}
\end{figure}

As a case study we chose the directions and injection points to represent some common case scenarios of down-going single particles. For each injected particle, we record the total number of PE detected by each PMT, as well as the time of the first photon detected on the photocathode. The directions and the entry positions of the particles in the WCD unit are shown in the left column of Fig. \ref{fig:Augerlike}, while the pictures on the right illustrate the resulting average PE distribution for each multiPMT channel of 1000 muons for each direction. In each case, the PE distribution shows evident non-uniformity and asymmetry across the PMTs.
Fig. \ref{fig:centroid} shows the spread of the charge centroids reconstructed from the signals of all the PMTs in the module for the different simulated events.
Further considerations can be made upon the temporal information, as the first photon interaction time stamps record the development of the Cherenkov light cone as it approaches the front-most PMTs first and the oppositely positioned PMTs last. We underline that these results are obtained by employing only the integrated charges and that future improvements could benefit from the inclusion of time evolution of the signals as captured by a flash ADC.


The results presented, performed at a fundamental level, aim at illustrating the wealth of information and the potential improvements obtained by adopting a direction-sensitive detector like a multiPMT as opposed to a single large PMT configuration in WCD units.
We find that the multiPMT shows sensitivity to the features of the light distribution originating from the Cherenkov light cones of single muons.
The potential of using multiple photosensors to determine the probability of muon presence in the WCD signal has already been explored with success.
In particular, it has been demonstrated that machine-learning algorithms can be trained to identify the presence of a muon in
the WCD unit. It has also been shown that the sum of such probabilities is directly proportional to the number of muons in the shower, paving the way for detailed composition studies of cosmic rays and contributing to gamma-hadron discrimination by selecting muon-poor showers as candidates for gamma-ray initiated events \cite{LIP, gonzalez2021tacklingmuonidentificationwater, Assis_2022}. The next section explores the feasibility of employing the multiPMT module in a WCD unit for muon tagging applications.

\section{Performance with Extensive Air Showers}
\begin{figure}
    \centering
    \includegraphics[width=0.48\linewidth]{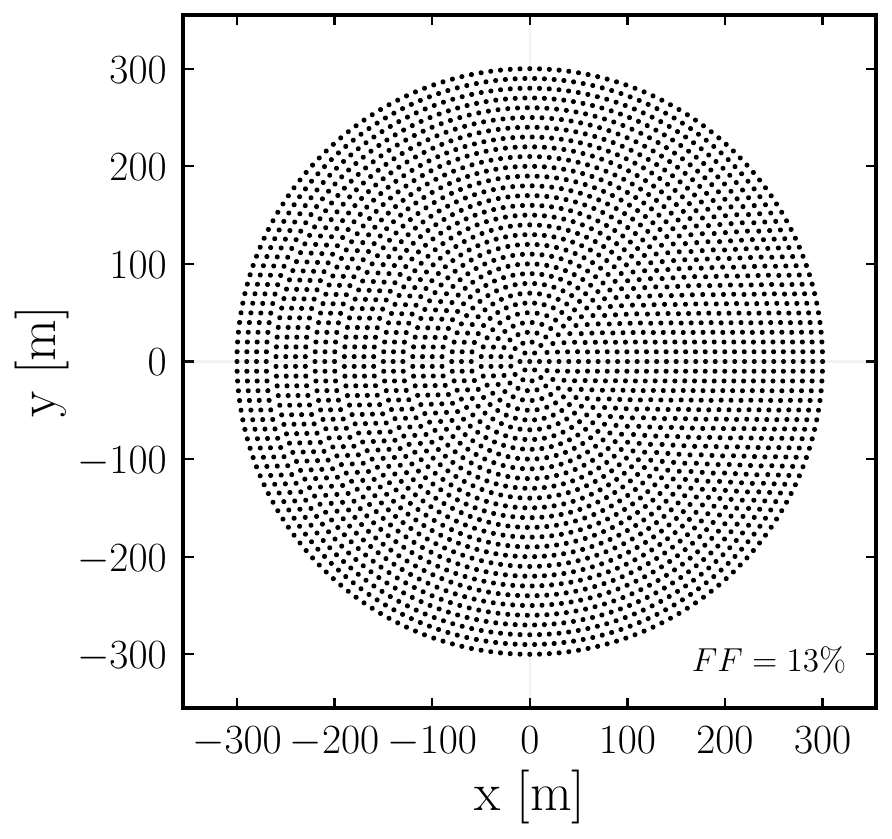}
    \caption{Sketch of the toy-model array of $13$$\%$ $\mathrm{FF}$ and radius $\mathrm{R} = 300$ $m$. }
    \label{fig:13FF_array}
\end{figure}

In order to assess the instrument performance in realistic conditions, we employ the WCD unit equipped with a multiPMT as described in Sec. \ref{sec:conceptual-module-description} to define a WCD array as a toy-model detector for sampling, at ground-level, CORSIKA-simulated \cite{Heck:1998vt} showers initiated by VHE protons. The choice of using an array layout as opposed to individual WCD units is dictated by the necessity of probing the photosensor response over various distances of the said units from the shower core. 
We consider WCD units equally spaced with a radial and on-circumference distance of $10$ $\mathrm{m}$ in
a circular array design of radius $\mathrm{R} = 300$ m as sketched in Fig. \ref{fig:13FF_array}. The size of the array matches the lateral extension of the EAS in the considered energy range and altitude. The resulting Fill Factor ($\mathrm{FF}$) is $\simeq 13$$\%$.

The CORSIKA simulations are performed assuming an observation altitude of 4700 m above sea level. A total of 62 proton-induced air showers are considered, with zenith angles in the range $0^{\circ}\leq\mathrm{\theta_{p}}\leq30^{\circ}$, primary energies between $100\leq \mathrm{E_{p}}\leq 110 $ TeV (sampled with an $E^{-2}$ spectrum) and with positions of the shower core in the center of the array. This configuration allows a certain degree of generalization in the momenta of the particles traversing the detector station.

The sampling strategy of the proton cascades at ground is performed as follows: for each WCD unit position in the toy-model array, we consider the batch of secondary particles falling at ground level within a radius of amplitude $R_{T} + r$ with $R_{T}$ being the radius of the tank and $r$ an extension that allows particles to enter laterally in the tank. The particles arrival time is computed with respect to a trigger time defined as the minimum arrival time of particles falling within $10$ m from the shower core. However, due to the fairly wide range of $\theta_{p}$ we introduce a correction to the particles arrival time based on simple geometric considerations. 
The correction returns the particle arrival times in the shower plane reference system, removing the geometrical delay due to the inclination of the shower front. Each WCD unit is simulated using \textsc{Geant4} by injecting batches of secondary particles hitting the WCD unit. In order to show the potentiality of the multiPMT for muon tagging applications, we distinguish all events in two sets: particle batches with only the electromagnetic shower component of secondary $\gamma$ and $e^\pm$ and events with only single muons, the latter obtained by removing all other particles from the batch. Lastly, we extract the number of $\mathrm{PE}$ within a time window of $100$ $\mathrm{ns}$ for each of the seven PMTs. We decided to focus on the class of pure single-muon events, as it has been shown to be highly effective for machine-learning–based muon tagging analyses \cite{LIP, Assis_2022, gonzalez2021tacklingmuonidentificationwater}.

In order to show that the choice of focusing on the single muons events, as opposed to the realistic case of the simultaneous presence of muons and $\gamma$ and/or electrons, is a meaningful case study, we first establish a detection threshold of 13 PEs (discussed in the following) and then define the fraction of muon-originated photoelectrons as:

\begin{equation*}
f^{\mu}_{PE} = \dfrac{N_{PE}^{\mu}}{N_{PE}^{tot}},
\end{equation*}
where $N_{PE}^{\mu}$ is the number of $\mathrm{PE}$ originated from muons and $N_{PE}^{tot}$ is the total number of $\mathrm{PE}$ in the event, which is an indicator of the purity of the event in terms of its compatibility with the single-muon case.
\begin{figure}[t!]
    \centering
    \includegraphics[width=0.48\textwidth]{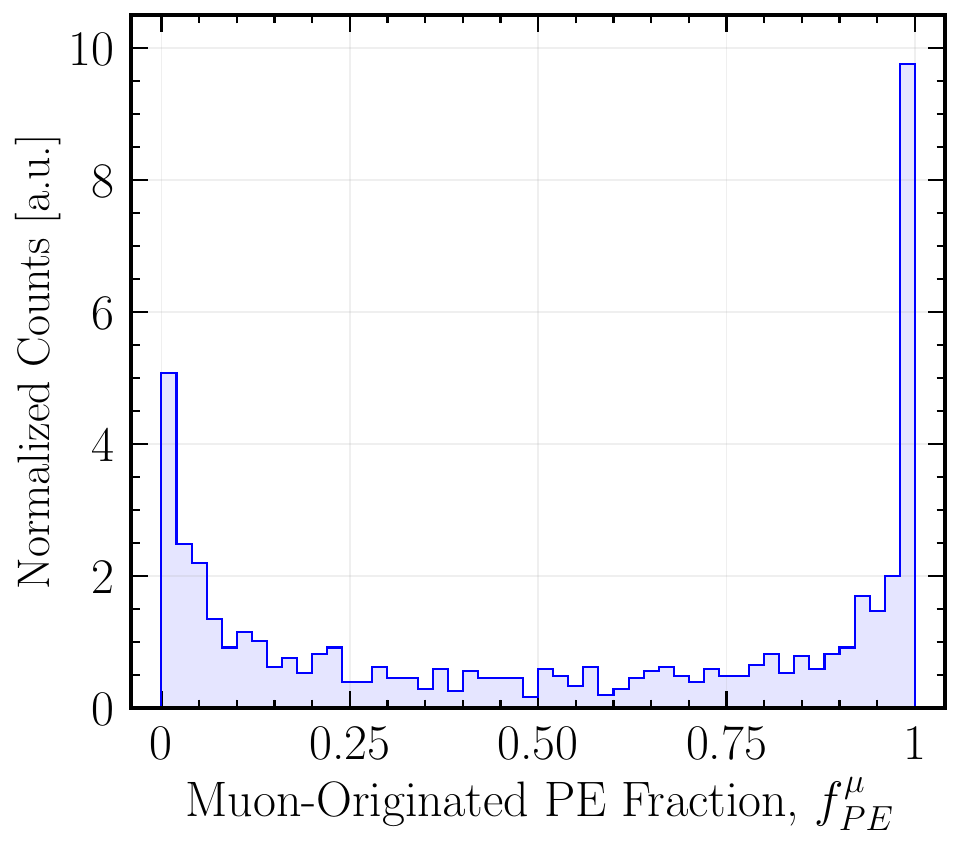}
    \hspace{0.3cm}
    \includegraphics[width=0.48\textwidth]{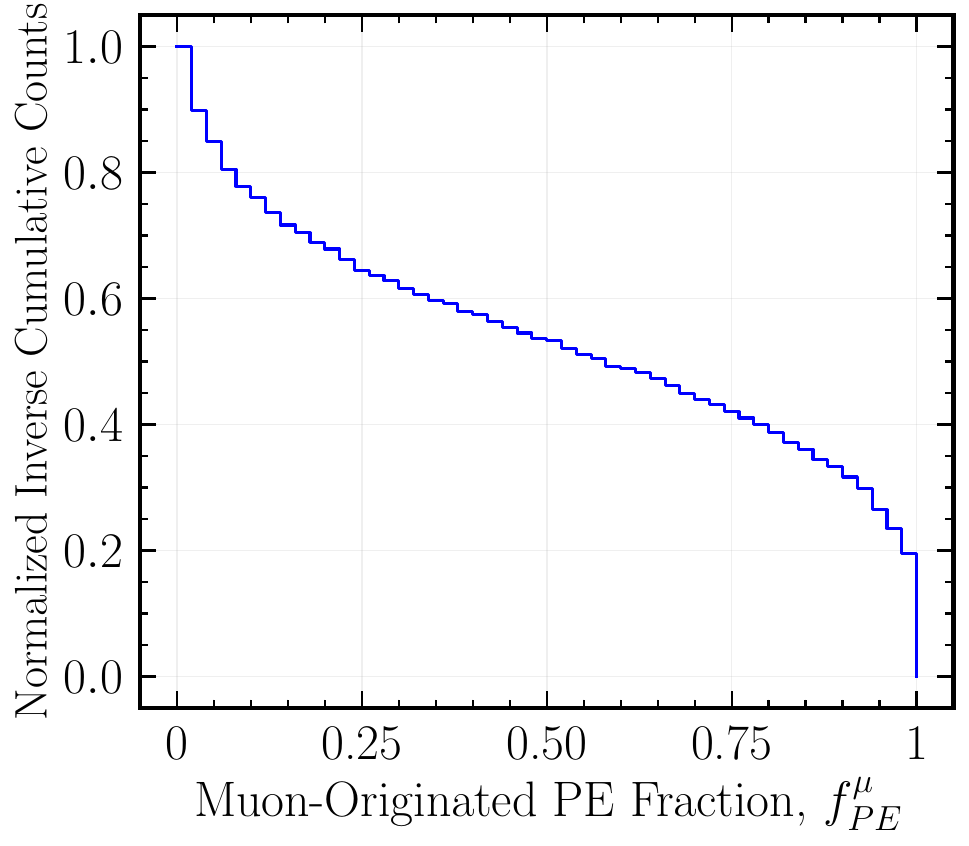}
    \caption{Distribution of the fraction of muon-originated photoelectrons in the general scenario of particle batches containing at least one muon (left) and inverse cumulative distribution (right). All events considered satisfy the requirement discussed below of having a total $\mathrm{PE}$ count greater than 13.}
    \label{fig:frac_mu_13pe}
\end{figure}
The probability distribution of the fraction of muon-originated photoelectrons and its inverse cumulative for our toy-model array show a significant set of muons with either minimal or no contamination at all from other particles that can be considered as single muon events, as shown in Fig. \ref{fig:frac_mu_13pe} .

\begin{figure}[t!]
	\centering
    \includegraphics[width=1\linewidth]{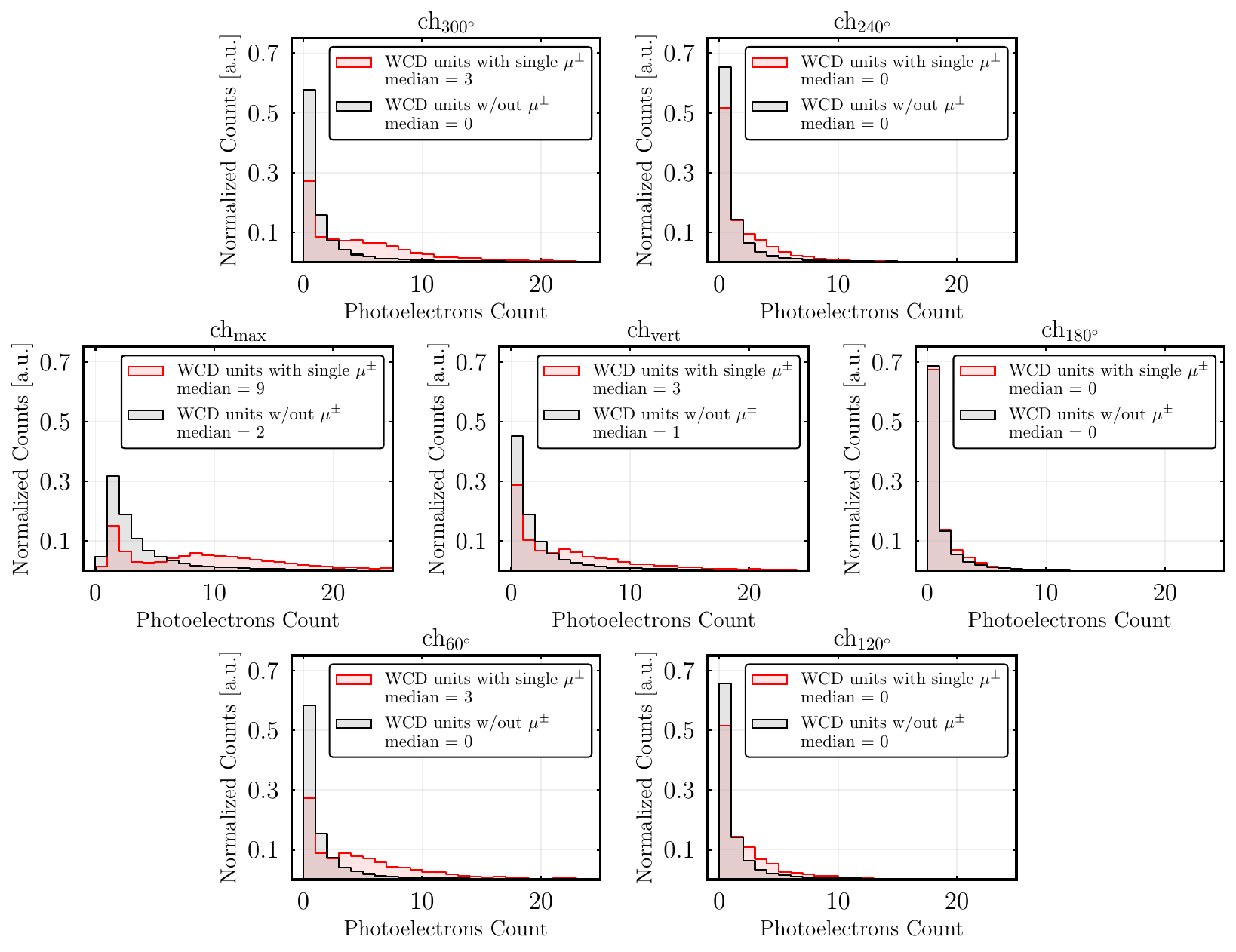}
	\caption{PE distributions for each of the seven PMT channels. The plots shown overlay the distributions for single muons (red) and muon-less events (black). Channels are labeled according to the convention described in the text. All simulated events shown have a minimum of $1$ total $\mathrm{PE}$ across all channels combined.  }
	\label{fig:mPMT_pe_per_channel_mu_nomu}
\end{figure}

Given the intrinsic azimuthal symmetry of the multiPMT, we introduce a labeling convention: we consider as a reference channel the lateral PMT registering the highest number of PE, denoting it as $\mathrm{ch}_{\text{max}}$, and label the remaining five lateral PMTs counterclockwise. The central upward-facing PMT is labeled $\mathrm{ch}_{\mathrm{vert}}$, yielding the set:  
{\large
\[
\mathrm{ch}_{\mathrm{vert}}, \; \mathrm{ch}_{\text{max}}, \; \mathrm{ch}_{60^{\circ}}, \;
\mathrm{ch}_{120^{\circ}}, \; \mathrm{ch}_{180^{\circ}}, \;
\mathrm{ch}_{240^{\circ}}, \; \mathrm{ch}_{300^{\circ}}
\]
}
with $\mathrm{ch}_{60^{\circ}}$ and $\mathrm{ch}_{300^{\circ}}$ naturally adjacent to $\mathrm{ch}_{\text{max}}$ and $\mathrm{ch}_{180^\circ}$ as its opposite-facing channel. This reference frame is particularly suited for studies on the asymmetry of the $\mathrm{PE}$ across PMTs. A time-based labeling of the channels could emphasize other aspects of the recorded event. In this paper, we focus on the presented reference frame convention, and postpone to future analyses an investigation on the additional temporal information content provided by a multiPMT. 
In Fig. \ref{fig:mPMT_pe_per_channel_mu_nomu} we show the $\mathrm{PE}$ distribution for each of the PMTs in the described labeling convention for all batches of secondary particles.

$\mathrm{ch}_{\max}$ and its adjacent channels ($\mathrm{ch}_{60^{\circ}}$, $\mathrm{ch}_{300^{\circ}}$) show a higher number of events with large PE counts and a higher median value, while opposite-side channels ($\mathrm{ch}_{180^{\circ}}, \mathrm{ch}_{240^{\circ}}, \mathrm{ch}_{300^{\circ}}$) are more populated at low PE counts with at least half of the total events contained in the first bin. This asymmetry is further enhanced when we analyze events containing only a single muon and no other particle. 
This is due to the large amount of light in a single Cherenkov cone produced by a single muon as compared to the more dispersed illumination from multiple Cherenkov cones produced by the more abundant $\gamma$ and $e^\pm$.
The plots in Fig.~\ref{fig:mPMT_pe_per_channel_mu_nomu} motivate the definition of a quantity to describe the PE asymmetry to trace the uneven light distribution typically produced by a muon in the WCD unit. Many expressions for asymmetry can be defined, in this work we propose the following definition of asymmetry based only on the six lateral channels, leaving other possible definitions for future studies.

\begin{equation} \large
A_{6} = \dfrac{(\mathrm{ch}_{\max}^{\mathrm{PE}} + \mathrm{ch}_{60^\circ}^{\mathrm{PE}} + \mathrm{ch}_{300^\circ}^{\mathrm{PE}})
-
(\mathrm{ch}_{180^\circ}^{\mathrm{PE}} + \mathrm{ch}_{120^\circ}^{\mathrm{PE}} + \mathrm{ch}_{240^\circ}^{\mathrm{PE}})
}{
(\mathrm{ch}_{\max}^{\mathrm{PE}} + \mathrm{ch}_{60^\circ}^{\mathrm{PE}} + \mathrm{ch}_{300^\circ}^{\mathrm{PE}})
+
(\mathrm{ch}_{180^\circ}^{\mathrm{PE}} + \mathrm{ch}_{120^\circ}^{\mathrm{PE}} + \mathrm{ch}_{240^\circ}^{\mathrm{PE}})
}
\label{eq:a6}
\end{equation}

\begin{figure}[t!]
    \centering
    \includegraphics[width=0.48\textwidth]{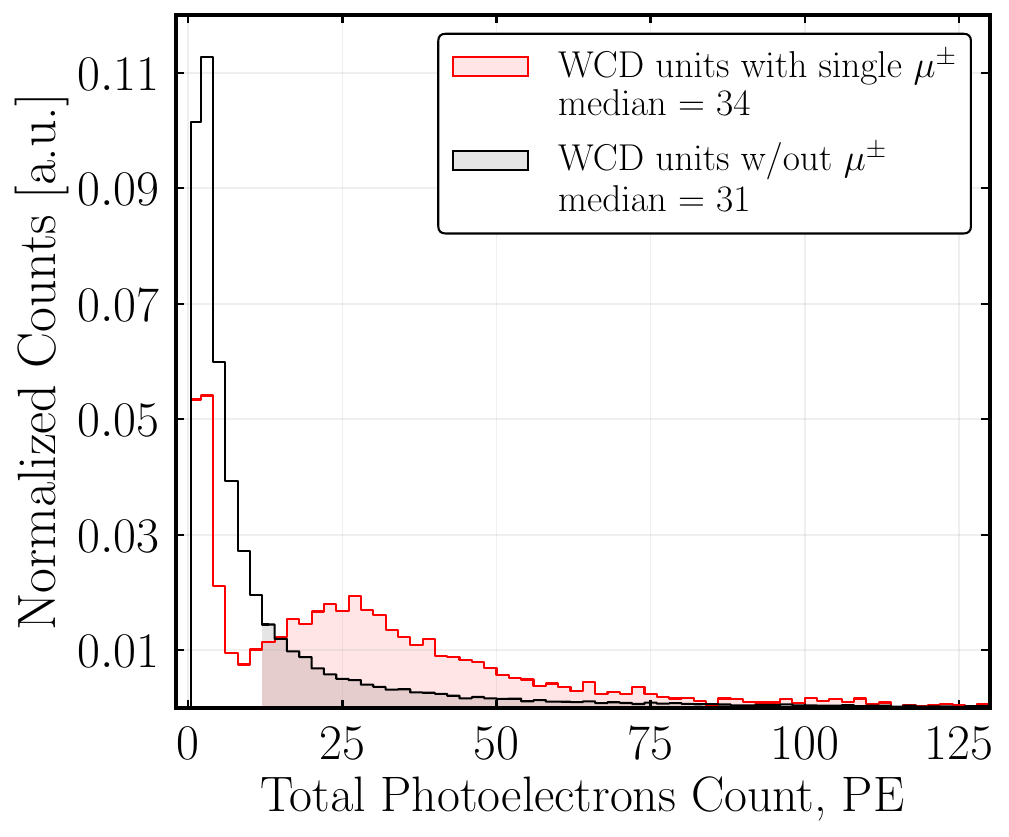}
    \caption{Distribution of the total number of $\mathrm{PE}$ across all channels for single muons (red continuous line) and muon-less events (black continuous line). The color-filled regions indicate events above the $13$ $\mathrm{PE}$ mark. The median values refer to the events satisfying the required condition.}
    \label{fig: mPMT_total_pe_above_13pe}
\end{figure}

With the superscript indicating the number of $\mathrm{PE}$ for the specific channel. Such defined asymmetry variable is restricted to the interval $-1 \le A_{6} \le 1$, where a higher asymmetric $\mathrm{PE}$ disposition across front-facing PMTs and opposite side PMTs would result in a value of $A_{6}$ closer to $1$ and a more homogeneous distribution would lead towards $A_{6} \simeq 0$. In the single muons case the majority of the Cherenkov photons is recorded in the PMTs adjacent to $\mathrm{ch}_{\max}$ with consequently large values of $A_{6}$. On the contrary, the soft electromagnetic component tends to come in batches with a random spatial disposition of the particles falling in the WCD unit, resulting in a more uniform distribution of light on the different channels.
Negative values of $A_{6}$ are uncommon and in most cases are the consequence of events with an overall small number of total $\mathrm{PE}$. Indeed, the variable $A_{6}$ is most effective when a cut is performed on the total $\mathrm{PE}$ number in order to avoid unfavorable situations in which only a few $\mathrm{PE}$s can create the illusion of high asymmetry. This cut introduced is also coherent with common analysis requirements. We place this cut at $13$ $\mathrm{PE}$ based on the total $\mathrm{PE}$ distributions obtained by summing $\mathrm{PE}$ counts across all channels for WCD units with one muon and without muons, shown in Fig. \ref{fig: mPMT_total_pe_above_13pe}. 
\begin{figure}[h]
    \centering
    \includegraphics[width=0.48\textwidth]{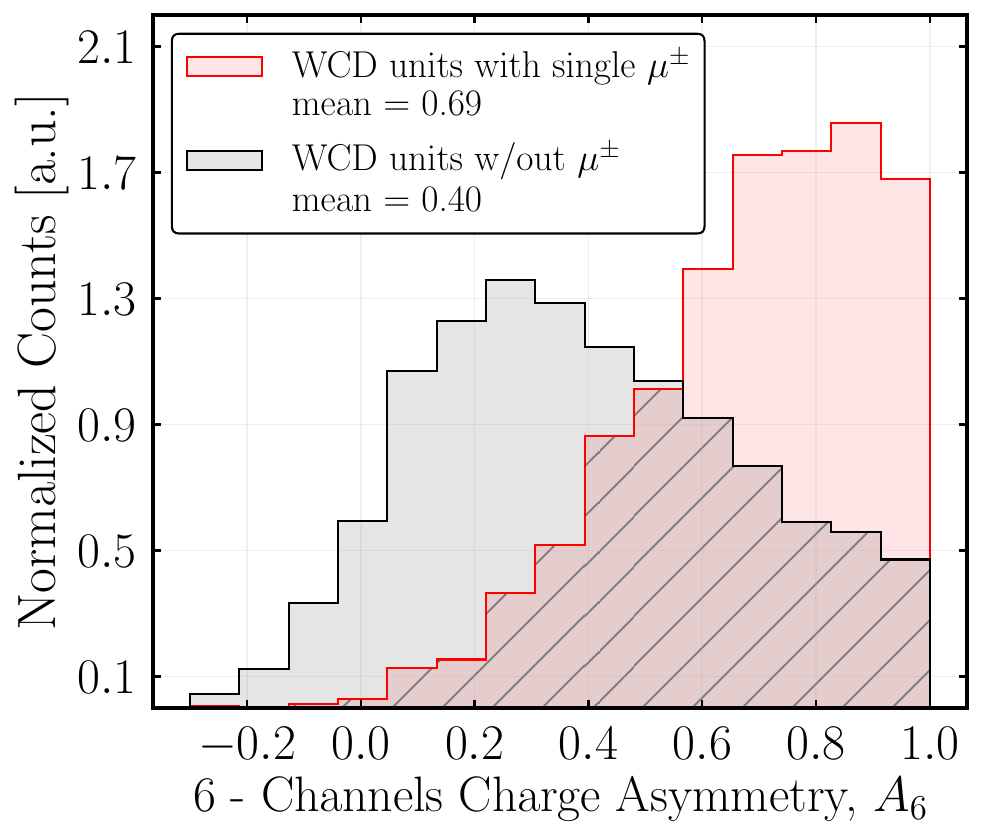}
    \caption{Distribution of the 6 - channels asymmetry variable $A_{6}$ as expressed in equation \ref{eq:a6} for WCD units traversed by a single muon (red) and for stations with $\gamma$ and $e^{\pm}$ and no muon (black).}
    \label{fig: asymmetry_6_channels}
\end{figure}

Finally, we show in Fig. \ref{fig: asymmetry_6_channels} the distributions of $A_{6}$ for single muons and muon-less WCD units with overall number of $\mathrm{PE} \ge 13$. We find, according to our expectation, that single muons behave more asymmetrically than batches of $\gamma$ and $e^{\pm}$; resulting in two distinct peaked distributions. This result is of particular interest when considering that, established a proper $\mathrm{PE}$ cut,
a single large PMT in the same configuration would be unable to further distinguish these classes of events.
\section{Conclusion}
This paper provides an overview of the usage of multiPMT modules in Water Cherenkov Detectors, highlighting their advantages and potential impact on high-energy astroparticle physics. We have demonstrated the potentiality of integrating a multiPMT system in small WCD units by studying the response of the system in a controlled single-station environment first and then using a realistic toy-model array.
We have shown that a multiPMT direction-sensitive design is able to grasp the different levels of asymmetry from muons and non-muon events through the photoelectron count distribution among its PMTs.
Ongoing research on WCDs continues to demonstrate the flexibility and effectiveness of the usage of multiple PMTs configurations. This paper opens the possibility of applying a multiPMT module in WCD units for muon tagging.

\bibliographystyle{ieeetr}
\bibliography{multiPMT.bib} 

\end{document}